# Event-Driven LSTM For Forex Price Prediction

Ling Qi
School of Computer Science
The University of Sydney
Sydney, Australia
liqi6811@uni.sydney.edu.au

Matloob Khushi
School of Computer Science
The University of Sydney
Sydney, Australia
matloob.khushi@sydney.edu.au

Josiah Poon
School of Computer Science
The University of Sydney
Sydney, Australia
josiah.poon@sydney.edu.au

*Abstract*— The majority of studies in the field of AI guided financial trading focus on purely applying machine learning algorithms to continuous historical price and technical analysis data. However, due to non-stationary and high volatile nature of Forex market most algorithm fail when put into real practice. We developed novel event-driven features which indicate a change of trend in direction. We then build long deep learning models to predict a retracement point providing a perfect entry point to gain maximum profit. We use a simple recurrent neural network (RNN) as our baseline model and compared with short-term memory (LSTM), bidirectional long short-term memory (BiLSTM) and gated recurrent unit (GRU). Our experiment results show that the proposed event-driven feature selection together with the proposed models can form a robust prediction system which supports accurate trading strategies with minimal risk. Our best model on 15-minutes interval data for the EUR/GBP currency achieved RME 0.006x $10^{-3}$, RMSE 2.407x$10^{-3}$, MAE 1.708x$10^{-3}$, MAPE 0.194% outperforming previous studies.

## I. INTRODUCTION

Forex Trading (FX) is the largest financial market in the world, consisting of multiple international participants including professionals and individuals who invest and speculate for profit due to its nature of robust liquidity. An automated system which could predict correct entry and exit points will help investors generate considerable profit with minimum risk. In recent years, machine learning has been used by many academics to study the exchange rate market. [1] has provided a summary of research in this field from 2009 to 2015.

As a type of non-stationary time-series data, financial trading data is highly volatile and complex. Technical analysis can smooth out noise and help to identify trends and has now become more popular in trading research. In addition, based on the "history repeating itself" theory, analysts believe patterns underlying historical data will repeat again in the future and being able to identify historical price movements is important for future price trend predictions. Therefore, the use of historical data and technical indicators are required for effective examination of trends that occur within desired trading windows. Literature has shown that a vast majority of work in this field is through historical data and technical analysis. This can be seen in [2] where a change of direction in the FX market was predicted by simply using the closing price and moving average individually with the moving average outperforming the closing price. More examples can be found from [3], [4], [5], [6], [7].

In this paper we will use an alternate method to handle the noisy and chaotic environment of high frequent trading data. In 1935, R.N. Elliott introduced "Elliott Wave Theory" [8]. One of its main components is to identify different waves in financial data. This includes impulse waves that set up a pattern, and corrective waves that oppose the larger trend. The identification of such waves helps us to discover the correct entry and exit trading points which would ultimately generate the most profit. Fig. 1 is an example of a typical Elliott Wave. We can see either peak or trough point "e1" and retrace point "e3" are the best market entry points and "e4" is the best market exit point.

The determination of peak or trough points within a certain window requires future price information. We therefore introduce a moving average crossover event ("e2"), as confirmation of the form of point "e1" and consider point "e3" as our prediction target to enter the market. Point "e4" can also be another prediction target to exit the market however we will not be covering that in this paper. Unlike other research that uses all historical data, we select data at "e2" and go back "n" timesteps as training data, while ignoring any other data which does not suggest a reliable trading opportunity. The selected data is then compressed with more relevant information.

We developed a LSTM model to predict price at the retracement point "e3". Technical indicators are used as features. We used four pairs of currency for experimentation. A simple RNN is used as the baseline model. The result confirms the designed architecture is over performing.

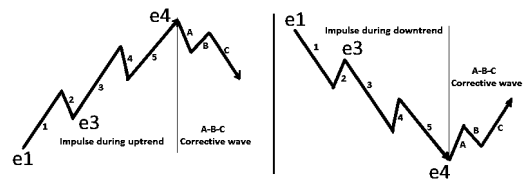

Fig.1 Example of Elliott Waves

(left is an upper trend, right is a downtrend)

The remainder of this paper is structured as follows: Section II discusses the related work. Section III describes details of feature and training data selection. Section IV explains model architecture, followed by experiment details in Section V. Our

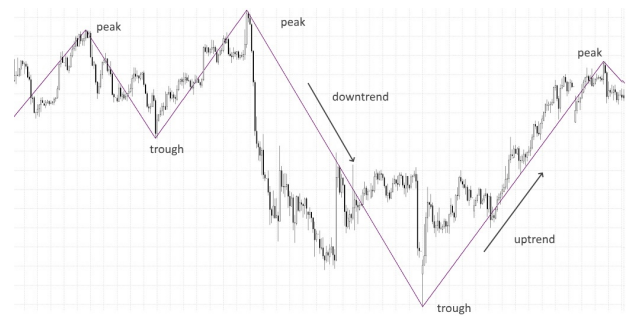

Fig.2 example of ZigZag

results appear in Section VI and Section VII presents our conclusion.

## II. RELATED WORK

Forecasts on FX trends have become challenging due to random fluctuations in price caused by market uncertainties such as political conditions, regulatory policy changes, economic conditions, banking operations and capital movements. Researchers have therefore tried a few varied approaches in this field. In order to gauge the direction of our study and the suitability of our own algorithm in generating reliable and valid results, we compare and discuss recent work in market prediction.

*A. Technical analysis indicators*

Technical analysis (TA) indicators are commonly used features in Forex prediction and many papers can confirm the effectiveness of TA indicators in forex forecasts. [9] analysed the performance of six simple machine learning models to predict a binary classification for price movements (up and down) for the USD/JPY currency pair. Nine features were generated from raw data based on technical indicators (MA, RSI and WR). [10] introduced a SVM to predict future price trend directions, it generated seventy indicators as features derived from technical analysis. The proposed model achieved 81% accuracy rate in forecasting future price trends.

Due to the rapid growth of computational power in recent years, a wide range of Neural Networks have been implemented in forex trading prediction. However, most papers conducting neural network research are still using historical price data and technical indicators as features for training. In 2011, [11] forecasts FX markets by feeding price and technical indicators into a neural network system. Similarly, [12] predicted future prices by using popular technical indicators – RSI, CCI, MACD and ROC. This paper focused more on building a trading agent to feed predicted prices into a decision module. Experiment results showed model effectiveness. [13] promoted a Convolutional Neural Network (CNN) model to forecast monthly and weekly price trends. The data features they chose were still technical indicators, however, they were used in conjunction with exchange rates, commodity prices and world indices. [13] achieved a 65% accuracy rate for monthly price trends and 60% for weekly price trends. [14] used nine indicators each with two different parameters together with a close price line chart as input for the CNN models. [15] used moving average 5 (MA5), moving average 10 (MA10), and moving average 20 information (MA20) line charts as input images to build a CNN to predict weekly price movements. The target movements were classified as up, down and non-movement based on movement percentage threshold. More examples can be seen from [16], [17], [18].

*B. Long Short-Term Memory (LSTM)*

During recent years, LSTM has gained popularity in time series prediction due to its capability to remember the previous inputs and its ability to prevent information from the past being lost. [19] introduced a feature fusion LSTM-CNN model for forecasting stock prices by taking the characteristics of both the chart image data and the time series data. The proposed model firstly used a SC-CNN model to extract hidden patterns in stock chart images, then used a ST-LSTM model to work on close prices and trading volumes. Kim and Kim's study tested performances on four different stock chart images and proved that candlestick charts are the best stock chart images to predict stock prices among bar charts, line charts and filled-line charts. This study also could not prevent the lagging issue that occurred in other LSTM stock price predictions. The authors then suggested to add noise-cancelling methods such as autoencoder or wavelet transformation or add technical indicators to the image to achieve better performance. [20] had a similar approach to extract features from CNN then feed CNN outputs to a LSTM memory to predict stock prices. [16] introduced a LSTM-based agent to learn the temporal pattern in data and developed an automatic trading system based on historical price data and current market conditions. [21] is another example of using LSTM to predict stock prices.

LSTM and its variations have proven their effectiveness in time series forecast, they have dominated the financial time series forecasting domain in recent years. [22] survey shows above 80 publications in the last 5 years.

## III. FEATURE AND TRAINING DATA SELECTION

*A. A sequence of events*

As mentioned previously the sequence of events was derived from Elliott Wave to determine peaks and troughs of a uptrend or downtrend. We then incorporate a ZigZag indicator for technical analysis. Zigzag is a popular technical indicator which identifies peaks and troughs by identifying the highest high or the lowest low within a certain period. Zigzag allows traders to observe price movements holistically and avoid market noise from small movements. ZigZag (e1) is the first event of the sequence.

Zigzag indicators can be modified or defined using three parameters, the Depth, Deviation and Backstep. The "Depth" is an integer which requires that a candidate Peak or Trough cannot have a lower low or higher high within the "Depth" range of the candidate period. "Deviation" refers to the amount of deviation (in pips) that is required to identify a new peak (from a trough) or a new trough (from a peak). Lastly, "Backstep" refers to the minimum number of periods required between adjacent peaks and troughs. Fig. 2 is an example of a ZigZag line.

ZigZag indicators are transition points in a certain time window. It could be the perfect entry or exit trading point to make a profit. However, the calculation of ZigZag requires future price information [19], thus we introduced the second event which can confirm the ZigZag point.

The second event we detected is a moving average crossover event (e2). Moving average is a calculation which produces a series of averages of combined price points of an instrument over a specified time frame. There are two commonly used moving averages - the simple moving average (SMA) and the exponential moving average (EMA). EMA is a weighted average of the last n prices, where the weighting decreases exponentially with each previous price/period. EMA gives a greater weight to more recent prices. Its formula is as follows:

$$EMA = \text{Price}(t) \times k + EMA(y) \times (1 - k)$$

where t is today, y is yesterday, N is number of days in EMA, and $k = \frac{2}{N+1}$.

A moving average crossover occurs when two or more moving average lines cross over each other. Amongst traders this is interpreted as a signal that a change in trend is occurring. In trading, the crossover point is often used as a trigger for a trading action, either to enter (buy or sell) or exit (sell or buy) in the market.

We lastly detected a retracement point (e3) which is also our prediction target. A retracement point is any temporary reversal in price within a major price trend. Retracement can be a confirmation of a trend. It can also help traders identify if the current trend is likely to continue or if a reversal is taking place. The right retracement point is the location for traders to enter into the market giving them the potential for good profits at minimum risk. In this paper we aim to predict price at the retracement point.

An example of a sequence of the above three events is shown in Fig. 3.

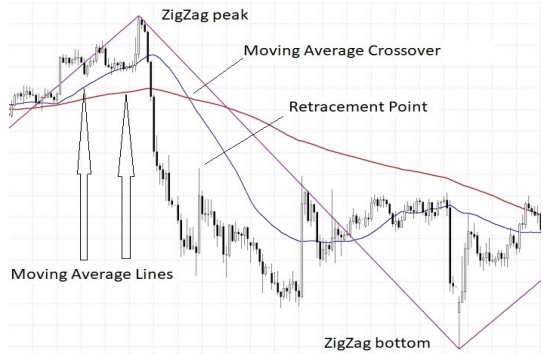

Fig.3 Example of a sequence of events

### B. Feature extraction

In addition to the above events, technical indicators are also used as features. A technical indicator is a mathematical calculation based on historical price, volume, or (in the case of futures contracts) open interest information that aims to forecast financial market direction. Technical indicators normally are used in conjunction with other techniques such as the occurrence of a sequence of events to determine the next trade action. We generated 28 technical indicators from a TA-LIB package for our models. The 28 indicators are derived from 6 types of indicators with different window size, namely Moving Average Convergence Divergence (MACD), Simple Moving Average (SMA), Relative Strength Index (RSI), Average Directional Index (ADX), Bollinger Band Indicator and William R Indicator (WR). All these indicators are non-volume based due to the difficulty in collecting reliable volume data.

### C. Training data selection

In order to capture price movements of time series data, most researchers chose whole continuous datasets as training data.

```
Algorighm
Origina Dataset X = { x_i | i = 0, 1, …… n }
1. Identify E1, E2, E3 for all data
2. Calculate 28 features
3. Loop through each E2
4.    go back n timesteps*
5.    generate 2D array for n timesteps * 28 features
6. Stack 2D array at all E2 as training data
7. Price at E3 is target to predict
* in this paper, we use 30 and 60 timesteps for experiment
```

Fig. 4 Algorithm for training data selection and process

To reduce noise from highly volatile frequent trading data, we selected all data at a crossover event (e2) and then go back n timesteps for training. Our prediction target is price at retracement point e3.

The feature process and training data selection method is described in Fig. 4.

## IV. MODEL ARCHITECTURE

RNN is a generalized feedforward neural network which is capable of processing sequences of data one element at a time while retaining an internal memory. RNN's recurrent nature performs the same function of every input of data and uses outputs from the previous input together with current timestep data as an input for functions. The recurrent nature has memory of the previous state and allows the network to learn long-term dependencies in a sequence and take the entire context into account when making a prediction. Fig. 5 depicts RNN architecture.

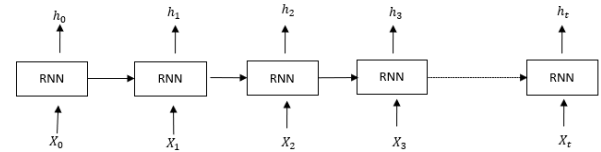

Fig. 5 Recurrent Neural Network Architecture

Although RNN has the capability to predict sequences of data, it has the disadvantage of gradient vanishing and exploding problems. LSTM is a modified version of RNN which resolves the vanishing gradient problem. LSTM architecture is composed of a cell and three gates – an input gate, an output

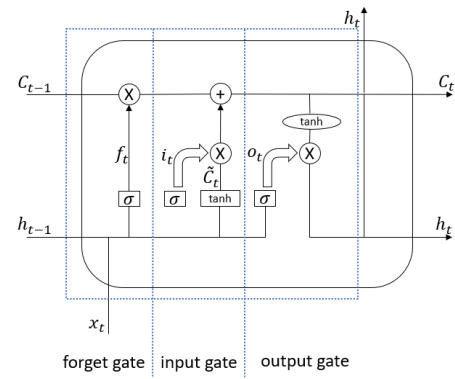

Fig. 6 Long Short-Term Memory Architecture

gate and a forget gate. Fig. 6 describes the LSTM architecture.

The forget gate investigates previous state $h_{t-1}$ and current input $x_t$ and determines which parts are to be discarded from the cell. The decision is made by a sigmoid function $f_t$ which outputs a number between 0 and 1 where 1 represents "keep" while 0 represents "remove". The sigmoid function formula can be denoted as below:

$$f_t = \sigma(W_f * [h_{t-1}, x_t] + b_f)$$

where $W_f$ represents weight function and $b_f$ represents bias.

The input gate decides which new information will be stored in the cell. A sigmoid function $i_t$ with formula as below will decide which values to be updated:

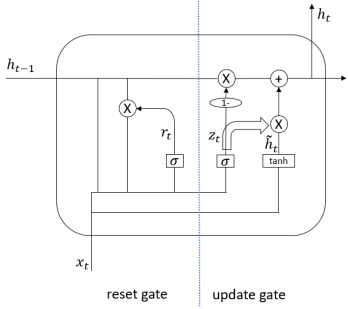

Fig. 7 Gated Recurrent Unit Architecture

$$i_t = \sigma(W_i * [h_{t-1}, x_t] + b_i)$$

Then, a tanh function creates a vector $\tilde{C}_t$ combined with $i_t$ to generate an update to the cell. $\tilde{C}_t$ can be denoted as below:

$$\tilde{C}_t = tanh(W_C * [h_{t-1}, x_t] + b_C)$$

The new cell $C_t$ is updated by information carried from the previous state and information generated from the current state, formula is as below:

$$C_t = f_t * C_{t-1} + i_t * \tilde{C}_t$$

Finally, the output gate will decide what to output. A sigmoid function $o_t$ will filter through the cell state to decide which parts to be sent to output. To generate output $h_t$, we then multiply $o_t$ by the cell state which comes through a tanh function to squish all values between -1 and 1. The details can be described as below:

$$o_t = \sigma(W_o * [h_{t-1}, x_t] + b_o)$$
$$h_t = o_t * \tanh(C_t)$$

BiLSTM is an upgrade version of LSTM. It runs inputs in two ways, one from past to future and one from future to past. The advantage of the approach is to preserve information from both past and future. BiLSTM architecture can be denoted as Fig. 8.

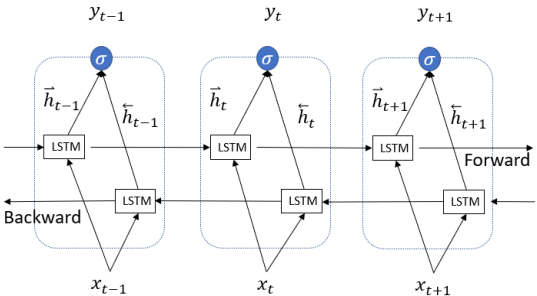

Fig. 8 Bi-directional LSTM Architecture

GRU is another variant of LSTM, the same as LSTM it has the capability to solve vanishing gradient problem comes with a standard RNN. The difference from LSTM is that GRU does not have an output gate and it has fewer parameters. GRU has two gates – an update gate and a reset gate which will decide what parts of information to be sent to output. The architecture can be denoted as Fig. 7.

In the reset gate, function $r_t$ decides how much past information to forget, its formula is as below:

$$r_t = \sigma(W_r * [h_{t-1}, x_t] + b_r)$$

The update gate then decides what information to remove through function $z_t$ and what new information to add via $\tilde{h}_t$, finally function $h_t$ generates the output.

$$z_t = \sigma(W_z * [h_{t-1}, x_t] + b_z)$$
$$\tilde{h}_t = tanh(W_h * [h_{t-1}, x_t] + b_h)$$
$$h_t = (1 - z_t) * h_t + z_t * \tilde{h}_t$$

GRU is faster to train than LSTM and more efficient. However, in general, LSTM performs better than GRU where long-term memory is required.

In this paper, the proposed LSTM is composed of an input layer, 2 LSTM layers with 64 hidden units on each layer together with a dense layer. Similarly, The BiLSTM is composed of an input layer, 2 BiLSTM layers with 64 hidden units and a dense layer. GRU has similar structure as LSTM with one input layer, 2 recurrent layers and 64 hidden units. All three models are trained with Adam optimizer [23].

These three models inform us when is the correct time to enter the market by forecasting price at the retracement point ("e3"). As we already know whether the market is a bullish or bearish market from event "e1" and "e2", the goal of the LSTM model is to decide at what price we should enter the market. Such a decision can be interpreted as a "buy" order in a bullish market or a "sell" order in a bearish market.

The whole process of the designed methodology is shown in Fig. 9.

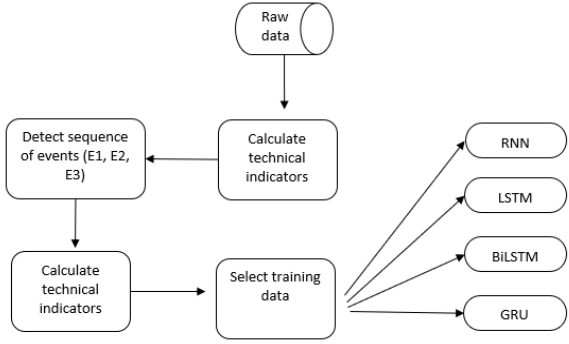

Fig. 9 Process of prediction system

## V. EXPERIMENTS

### A. Data and Features

In this paper we use four major currency pairings – GBP/USD, EUR/GBP, AUD/USD and CAD/CHF with their 15 minutes interval data from Jan 2005 to Sep 2017 for training and data from Oct 2017 to Sep 2020 for testing. Data was extracted from Oanda API. The total number of sequences of events for each currency pair are listed in table I.

The 28 features are MACD, MACD signal, MACD histogram, SMA5, SMA10, SMA15, SMA20, SMA25, SMA30, SMA36, RSI5, RSI14, RSI20, RSI25, ADX5, ADX10, ADX15, ADX20, ADX25, ADX30, ADX35, Bollinger lowerband, Bollinger middleband, Bollinger upperband, WR5, WR14, WR20 and WR25.

TABLE I. RAW DATA DETAIL

| Currency Pair | Number of rows | Number of sequences (e1, e2, e3) | Training data | Test data |
|---|---|---|---|---|
| GBP/USD | 397,317 | 2,238 | 1,838 | 400 |
| EUR/GBP | 397,796 | 2,233 | 1,810 | 423 |
| AUD/USD | 397,329 | 2,260 | 1,849 | 411 |
| CAD/CHF | 397,813 | 2,194 | 1,770 | 424 |

### B. Evaluation Measures

The performance of the model was measured by standard regression metrics including mean square error (MSE), root mean squared error (RMSE), mean absolute error (MAE) and mean absolute percentage error (MAPE).

The MSE measures the average of the squares of the errors, its formula is as follows:

$$MSE = \frac{\sum_{i=1}^{n}(True - Prediction)^2}{n}$$

The RMSE is the root of MSE, its formula is as follows:

$$RMSE = \sqrt{\frac{\sum_{i=1}^{n}(True - Prediction)^2}{n}}$$

The MAE is the average of absolute error. It measures errors between paired observations. Its formula is as follows:

$$MAE = \frac{\sum_{i=1}^{n}|True - Prediction|}{n}$$

The MAPE measures the size of the error in percentage terms. It is calculated as the mean of the absolute percentage errors of forecasts. Its formula is as follows:

$$MAPE = \frac{\sum_{i=1}^{n}\frac{|True - Prediction|}{True}}{n} * 100$$

## VI. RESULTS

Our experiment results are shown in Table II, Table III, Table IV and Table V. The results show that LSTM with 30 timesteps is the best performing model for EUR/GBP, GRU with 30 timesteps is the best model for both AUD/USD and CAD/CHF. These three models are all over-performing against the baseline RNN model with the exception of GBP/USD for which RNN with 60 timesteps is the best model.

Fig. 10 shows the comparison of the ground truth (i.e. real price at retracement point) and predicted price for EUR/GBP. The predicted values are very close to the true values and its MAPE is only 0.194% which gives us confidence that the model can support real trading.

## VII. DISCUSSION AND CONCLUSION

This research is aimed at predicting future price movements of Forex based on training data, features selected from historical data and technical analysis indicators utilising advanced AI techniques. We built an architecture constructed of a training data selection module and a LSTM model to predict price at retracement points. The forecast of an entry can help traders to determine their trading strategies with minimal risk.

While the results to date show great potential for a combined approach to develop a successful trading strategy, one of the main limitations is that we predicted "e1" using ZigZag indicators. Our ZigZag normally requires approximately an additional 40 bars to confirm a ZigZag transition point. We suggest further research on identifying ZigZag or similar price transition points based on historical data only.

TABLE II. EXPERIMENT RESULT FOR GBP/USD

| Model | Timesteps | MSE ($10^{-3}$) | RMSE ($10^{-3}$) | MAE ($10^{-3}$) | MAPE (%) |
|---|---|---|---|---|---|
| RNN | 30 | 1.846 | 42.960 | 35.336 | 2.774 |
| LSTM | 30 | 2.435 | 49.351 | 48.357 | 3.749 |
| BiLSTM | 30 | 1.520 | 38.984 | 37.456 | 2.903 |
| GRU | 30 | 6.311 | 79.441 | 78.185 | 6.067 |
| RNN | 60 | **0.710** | **26.645** | **25.118** | **1.945** |
| LSTM | 60 | 0.853 | 29.200 | 26.617 | 2.075 |
| BiLSTM | 60 | 1.489 | 38.593 | 37.098 | 2.873 |
| GRU | 60 | 3.324 | 57.658 | 56.723 | 4.401 |

TABLE III. EXPERIMENT RESULT FOR EUR/GBP

| Model | Timesteps | MSE ($10^{-3}$) | RMSE ($10^{-3}$) | MAE ($10^{-3}$) | MAPE (%) |
|---|---|---|---|---|---|
| RNN | 30 | 0.032 | 5.628 | 4.456 | 0.503 |
| LSTM | 30 | **0.006** | **2.407** | **1.708** | **0.194** |
| BiLSTM | 30 | 0.168 | 12.962 | 12.701 | 1.440 |
| GRU | 30 | 0.151 | 12.273 | 11.343 | 1.277 |
| RNN | 60 | 0.075 | 8.684 | 7.270 | 0.817 |
| LSTM | 60 | 0.006 | 2.536 | 1.957 | 0.222 |
| BiLSTM | 60 | 0.076 | 8.705 | 8.434 | 0.957 |
| GRU | 60 | 0.135 | 11.612 | 10.548 | 1.187 |

TABLE IV. EXPERIMENT RESULT FOR AUD/USD

| Model | Timesteps | MSE ($10^{-3}$) | RMSE ($10^{-3}$) | MAE ($10^{-3}$) | MAPE (%) |
|---|---|---|---|---|---|
| RNN | 30 | 0.530 | 23.030 | 20.406 | 2.875 |
| LSTM | 30 | 0.568 | 23.835 | 22.481 | 3.149 |
| BiLSTM | 30 | 4.821 | 69.433 | 68.646 | 9.566 |
| GRU | 30 | **0.201** | **14.162** | **13.588** | **1.929** |
| RNN | 60 | 1.061 | 32.578 | 31.106 | 4.377 |
| LSTM | 60 | 0.345 | 18.585 | 16.365 | 2.291 |
| BiLSTM | 60 | 6.603 | 81.263 | 80.897 | 11.320 |
| GRU | 60 | 0.298 | 17.270 | 16.626 | 2.351 |

TABLE V. EXPERIMENT RESULT FOR CAD/CHF

| Model | Timesteps | MSE ($10^{-3}$) | RMSE ($10^{-3}$) | MAE ($10^{-3}$) | MAPE (%) |
|---|---|---|---|---|---|
| RNN | 30 | 0.089 | 9.450 | 7.388 | 1.023 |
| LSTM | 30 | 0.092 | 9.570 | 8.537 | 1.167 |
| BiLSTM | 30 | 0.225 | 15.011 | 14.312 | 1.931 |
| GRU | 30 | **0.027** | **5.247** | **4.264** | **0.583** |
| RNN | 60 | 0.183 | 13.543 | 12.030 | 1.641 |
| LSTM | 60 | 0.154 | 12.396 | 11.088 | 1.514 |
| BiLSTM | 60 | 0.032 | 5.674 | 4.858 | 0.651 |
| GRU | 60 | 0.032 | 5.661 | 4.340 | 0.603 |

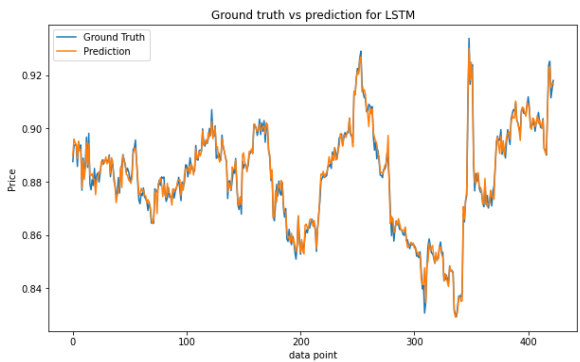

Fig. 10 Real Prices vs Predicted Prices – EUR/GBP

The main contributions of this paper are 1) We developed a method to select training data based on a sequence of events derived from Elliott Wave to reduce noise from frequent time series trading data. 2) Our LSTM and GRU models are outperforming the baseline RNN model for three currency pairs – EUR/GBP, AUD/USD and CAD/CHF. These results show a significant increase of MAPE - 159% for EUR/GBP, 49% for AUD/USD and 75% for CAD/CHF. 3) In addition to entry price point prediction, we tested and determined what was the best number of timesteps to use to capture an underlying price pattern. Results have shown currency pairings perform differently with different timesteps. Experiment results have also shown great potential to build a RoboTrading (Robotic Trading) platform.

Our results in this paper outperform previous results by Zeng and Khushi [24] they reported their lowest RMSE 1.65x10$^{-3}$ whereas our lowest RMSE in this paper is 0.006x10$^{-3}$.

ACKNOWLEDGMENT

Many thanks to Dr. Matloob Khushi and Dr. Saqib Ayyubi for their guidance and explanation of financial events. Also thanks to Ken Pang, Zhuoyang Li, YewShien Lee, Yuan Zeng and Jimmy Yue for sharing their knowledge.